\documentstyle[amsfonts,epsfig,12pt]{article}
\newcommand{\be}{\begin{equation}}
\newcommand{\ee}{\end{equation}}
\newcommand{\bea}{\begin{eqnarray}}
\newcommand{\eea}{\end{eqnarray}}
\newcommand{\beas}{\begin{eqnarray*}}
\newcommand{\eeas}{\end{eqnarray*}}
\newcommand{\bi}{\begin{itemize}}
\newcommand{\ei}{\end{itemize}}
\newcommand{\bc}{\begin{center}}
\newcommand{\ec}{\end{center}}
\newcommand{\bfl}{\begin{flushleft}}
\newcommand{\efl}{\end{flushleft}}
\newcommand{\bfr}{\begin{flushright}}
\newcommand{\efr}{\end{flushright}}
\newcommand{\f}{\frac}

\def\6{\partial} \def\a{\alpha} \def\b{\beta}
\def\g{\gamma}  
\def\e{\epsilon}
  
  \def\l{\lambda}
\def\m{\mu} \def\n{\nu}  
 \def\s{\sigma} \def\t{\tau}
  
\def\o{\omega} \def\G{\Gamma}

\newcommand{\HH}{{\cal H}}

\newcommand{\LL}{{\cal L}}

\newcommand{\OO}{{\cal O}}

\newcommand{\SS}{{\cal S}}

\newcommand{\ZZ}{{\cal Z}}
\begin{document}
\title{Thermal $D$-Brane Boundary States from Green-Schwarz Superstrings}
\author{{{Ion V. Vancea\footnote{ionvancea@ufrrj.br}}}}

\maketitle

\begin{center}
{{\em Departamento de F\'{\i}sica, Universidade Federal Rural do Rio de Janeiro},\\
{\em BR 465-07-Serop\'{e}dica CEP 23890-000 RJ, Brasil}}
\end{center}

\begin{abstract}
In this paper we thermalize the type II superstrings in the GS formulation by applying the TFD formalism. 
The thermal boundary conditions on the thermal Hilbert space are obtained from the BPS $D$-brane boundary conditions 
at zero temperature. We show that thermal boundary states can be obtained by thermalization from the BPS $D$-branes 
at zero temperature. These new states can be interpreted as thermal $D$-branes. Next, we discuss the supersymmetry 
breaking of the thermal string in the TFD approach. We identify the broken supersymmetry with the 
$\epsilon$-transformation while the $\eta$-transformation is preserved.  
Also, we compute the thermal partition function and the entropy of the thermal string.  
\end{abstract}

\newpage

\section{Introduction}

Recently, there has been an increasing interest in understanding the statistical properties of the $D$-branes motivated mainly by their phenomenological applications to the strong interactions, the physics of the early Universe and the cosmology, the thermodynamics of the black-holes and the thermal dualities of the string theories. While most of the results obtained thus far concern the calculation of the thermodynamical functions of the strings in the presence of the $D$-branes and the relations among the area, the entropy and the information in several $D$-brane models, less attention was given to the structural problem of the thermal $D$-branes.  In particular, one can address the question of the microscopic description of the $D$-branes at finite temperature and their relation to the thermal string states. This problem can be studied in those limits of the string theories in which there is an explicit formulation of the $D$-branes, like for example, in the perturbative string theory in the Minkowski spacetime where the $D$-branes are represented by boundary states constructed from the physical Hilbert space of the perturbative strings.

In previous works \cite{ivv1,ivv2,ivv3}, we proposed a new method for obtaining thermal boundary states from bosonic $D$-brane states (see for reviews and related problems \cite{ivv4,ivv5,ivv6}.) The main idea presented there is to thermalize the bosonic closed string by applying the Thermo Field Dynamics formalism (TFD) \cite{ubook}. From it, one can either derive the boundary conditions at finite temperature following the same procedure as at zero temperature or to thermalized the zero temperature boundary conditions. The corresponding thermal boundary states are coherent states in the Fock space of the thermal excitations of the string and can be interpreted as thermal $D$-branes. They describe the behavior of the excitations of the thermal bosonic closed string on the boundary. The statistical properties of strings and string field in the TFD framework have been previously studied in \cite{yl1,yl2,yl3,yl4,yl5,yl6,fn1,fn2,fn3,fn4,fn5,f1}. More recently, the analysis of thermal bosonic open string states \cite{ivv7} and the calculation of string entropy from TFD in $pp$-wave background were performed in \cite{ng1,ng2}. An important result was obtained in \cite{ng3} where it was shown the equivalence between the TFD and the Matsubara formalisms, respectively, in the string theory (see also \cite{ng4,ng5}.) Although the thermodynamical functions for ensembles of string excitations in the presence of $D$-branes can be derived in the Matsubara formalism too, in the TFD approach one can determine the actual form of the thermal boundary state from the states of the thermalized string in the same way as it is done at zero temperature. Also, the thermal vacuum state (and, in general, all thermal states) can be obtained and the symmetry breaking can be explicitely shown. Therefore, the method given in \cite{ivv1,ivv2,ivv3} can be considered a generalization of the boundary state formalism to finite temperature string theory. In this paper we are going to extend the method to the BPS $D$-branes of the type II strings.

The description of the BPS $D$-branes as boundary states can be given in the Green-Schwarz (GS) superstring theory in which the spacetime supersymmetry is manifest \cite{mg,gg,kh,lw}. In this formulation the non-physical degrees of freedom are eliminated by choosing the light-cone gauge. Consequently, the transverse degrees of freedom are classified according to the massless representations of the $SO(8)$ group. In particular, this approach is suitable for calculating the thermal boundary states from the BPS $D$-branes from the TFD since the quantization is canonical and the thermalization is automatically performed over the physical degrees of freedom only (see \cite{fns,fnn,ivv8} for attempts to apply the  TFD method to ghost fields.) Moreover, the linear independence of the string oscillators simplifies the structure of the thermal Hilbert space. 

This paper is organized as follows. In the next section we give a brief review of the construction of the BPS $D$-branes in the type II GS superstring. We thermalize the superstring and its boundary condition in Section 3. Here, we discuss the structure of the thermal vacuum and give the thermal boundary states. In Section 4 we analyse the broken supersymmetries of the thermal string and boundary states and discuss the finite temperature partition function and entropy of the thermal string. The last section is devoted to conclusions. Some brief review of TFD formalism and few formulas that are used throughout the text are collected in the appendices.

\section{Review of the BPS $D$-branes in GS Formalism}

In this section we are going to review the type II GS superstring and the construction of the BPS $D$-branes following \cite{gsw} and \cite{gg}. In the Minkowski target space the light-cone gauge is defined by taking $X^{+} = x^{+} + p^{+}\tau$ and solving for the coordinate $X^{-}$. The transverse bosonic coordinates of the superstring are denoted by $X^{I}$, where $I=1, 2 \ldots ,8$. They belong to the vector representation $\bold{{8}_{v}}$ of $SO(8)$. The fermionic coordinates $S^{a}$ and $\bar{S}^{a}$ are spinors from either $\bold{{8}_{s}}$ or $\bold{{8}_{c}}$ representations of $SO(8)$. The type IIA $D$-branes are constructed from superstrings of opposed left and right chiralities, while the type IIB $D$-branes are obtained from superstrings of the same chirality. The action for the type II superstring in the light-cone gauge is given by the following relation \cite{gsw}
\be
\SS_{l.c.g.} = -\frac{1}{2} \int d^2 \sigma \left(T \6_{\a}X^{I}\6^{\a}X^{I} - \frac{i}{\pi}\bar{S}^{*A} \Gamma^{-} \rho^{\a}_{AB}\6_{\a}S^{B}\right),
\label{stringaction}
\ee
where $S^{*Aa}=S^{\dagger Bb}(\G^{0})^{ba}(\rho^0)^{BA}$ represents the conjugation of a spinor in two and eight dimensions and $A, B = 1,2$. By $\rho^{\a}$'s we denote the complex Dirac matrices in two dimensions while $\G^{\m}$'s are the $32 \times 32$ imaginary Dirac matrices for the spinor representation of $SO(1,9)$ that satisfy $\{ \G^{\m}, \G^{\n} \} = -2\eta^{\m\n}$. Also, $\g^{I}$'s stand for the $16 \times 16$ real symmetric Dirac matrices for the spinor representation of $SO(8)$ that satisfy $\{ \g^{I},\g^{J} \} = 2\delta^{IJ}$. The last two sets of Dirac matrices are related by
\be
\G^{0}=\s_{2} \otimes 1_{16}~~,~~\G^{I}=i\s_{1} \otimes \g^{I}~~,~~\G^{9}=i\s_{3} \otimes 1_{16},
\label{Diracmatrices}
\ee
where $\s_i$ are the Pauli matrices in the standard representation. The upper off-diagonal elements of $\g^I$ form the matrix $\g^{I}_{a\dot{a}}$ that acts among the three representations of $SO(8)$. 

The action (\ref{stringaction}) has two supersymmetries. For example, in the type IIA superstring, the supercharges that generate the supersymmetries can be written as two inequivalent spinors $Q^{a}$ and $Q^{\dot{a}}$ from $\bold{{8}_{s}}$ and $\bold{{8}_{c}}$, respectively. The explicit form of the left-moving supercharges is given by the following relations
\be
Q^{a} = \frac{1}{\sqrt{2p^{+}}}\int^{\sigma}_{0} d\sigma \, S^{a}(\sigma )~,~
Q^{\dot{a}} = \frac{1}{\pi\sqrt{p^{+}}}\int^{\sigma}_{0} d\sigma \, \gamma^{I}_{\dot{a}b} \6 X^{I}  S^{b}(\sigma ).
\label{charges}
\ee
Similar relations hold for the right-moving supercharges. In the type IIB theory the two supercharges belong to the same representation. It is useful to write the superstring fields and the supercharges in terms of their Fourier transforms \cite{gsw}. The most general solutions of the equations of motion for the type IIB superstring have the form
\bea
X^{I}(\t ,\s ) &=& x^{I} + 2\a 'p^{I}\tau + i\sqrt{\frac{\a '}{2}}\sum_{n \neq 0}\frac{1}{n} \left( \a^{I}_n e^{-2in(\t -\s )} + \bar{\a}^{I}_n e^{-2in(\t + \s )} \right),\label{Fourierbos}\\
S^{1a}(\t ,\s ) &=& \sum_{n}S^{a}_n e^{-2in(\t -\s )}~~,~~
S^{2a}(\t ,\s ) = \sum_{n}\bar{S}^{a}_n e^{-2in(\t +\s )},\label{leftrightFourierferm}
\eea
where $\bar{~}$ denotes the right-moving modes. In the light-cone gauge the supersymmetry transformations of superstring coordinates can be written as two independent transformations that take the following form
\bea
\delta X^{I} &=& \sqrt{2/p^{+}}\left( \g^{I}_{a\dot{a}}\epsilon^{*\dot{a}}S^{a} +
\g^{I}_{a\dot{a}}\bar{\epsilon}^{*\dot{a}}\bar{S}^{a}\right), \label{susyX}\\
\delta S^{a} &=& \sqrt{2 p^{+}}\left( \eta^{a} - i \rho^{\a}\6_{\a}
{\cal X}^{I}\g^{I}_{a\dot{a}}\epsilon^{\dot{a}}\right),
\label{susyS}\\
\delta \bar{S}^{a} &=& \sqrt{2 p^{+}}\left( \bar{\eta}^{a} - i \rho^{\a}\6_{\a}
\bar{\cal X}^{I}\g^{I}_{a\dot{a}}\bar{\epsilon}^{\dot{a}}\right),
\label{susybarS}
\eea 
where $ X^{I} = {\cal X}^{I} + \bar{\cal X}^{I}$ is the decomposition of the bosonic field in to the right- and left-moving components, respectively. The $SO(8)$ spinors $(\eta, \bar{\eta})$ and $(\epsilon, \bar{\epsilon})$ parametrize the $\eta$- and $\epsilon$-supersymmetry of the right-moving and left-moving fields, respectively. The corresponding generators of the above transformations are 
\bea
Q^{a} &=& i\sqrt{p^{+}}(\G^{+}S_{0})^{a} + 2i\sqrt{1/p^{+}}\sum_{n=-\infty}^{+\infty}(\g_{I}S_{-n})^{a}\a^{I}_{n}
\label{susygenQ},\\
\bar{Q}^{a} &=& i\sqrt{p^{+}}(\G^{+}\bar{S}_{0})^{a} + 2i\sqrt{1/p^{+}}\sum_{n=-\infty}^{+\infty}(\g_{I}\bar{S}_{-n})^{a}\bar{\a}^{I}_{n}
\label{susygenbarQ}.
\eea
The BPS $Dp$-branes are defined by imposing $I=1,2,\ldots , p+1$ Neumann boundary conditions and $I=p+2,\ldots , 8$ Dirichlet boundary conditions on the superstring equations of motion and requiring that half of the supersymmetries of the theory be preserved. For definites, let us choose the type IIB superstring theory. We define the following linear combinations of supercharges
\be
Q^{\pm a}=(Q^{a}\pm iM_{ab}\bar{Q}^{b})~,~Q^{\pm \dot{a}}=(Q^{\dot{a}}\pm iM_{\dot{a}\dot{b}}\bar{Q}^{\dot{b}}).
\label{lincharge}
\ee
The charges $Q^{\pm a}$ and $Q^{\pm \dot{a}}$ contain informations about the presence of the hypersurface defined by the $Dp$-brane boundary conditions in the matrices $M$. Upon quantization, they become operators that act on the Hilbert space of the superstring. The breaking of only half of the supersymmetries by the BPS $Dp$-brane boundary state $|B\rangle$ reduces to the following set of constraints on the Hilbert space \cite{gg}
\bea
(\partial X^{I}-M_{J}^{I}\overline{\partial}X^{J})|B\rangle=0\nonumber,\\
Q^{+a}|B\rangle=Q^{+\dot{a}}|B\rangle=0\label{zeroTbc}.
\eea
Beside their dependence on the specific choice of Dirichlet and Neumann boundary conditions for the bosonic coordinates, the matrices $M^{I}_{J}$, $M_{ab}$ and $M_{\dot{a}\dot{b}}$ depend on the background fields, too. They have the following general form 
\be
M_{IJ}=\exp \{ \Omega_{KL}\Sigma^{KL}_{IJ} \}~,
M_{ab} = \exp \{  \frac{1}{2}\Omega_{IJ}\gamma^{IJ}_{ab} \}~,~
M_{\dot{a}\dot{b}} = \exp \{ \frac{1}{2}\Omega_{IJ}\gamma^{IJ}_{\dot{a}\dot{b}} \},
\label{fermMatrix}
\ee
where $\Omega \in SO(8)$ and 
\be
\Sigma^{KL}_{IJ} = \left( \delta^{K}_{~I}\delta^{L}_{~J} - \delta^{L}_{~I}\delta^{K}_{~J}\right)~,~
\gamma^{IJ}_{ab} = \frac{1}{2}(\gamma^I_{a\dot{a}}\gamma^J_{\dot{a}b} - \gamma^J_{a\dot{a}}\gamma^I_{\dot{a}b}).
\label{gamma}
\ee
In the absence of the open superstring condensates on the boundary the matrix $M_{IJ}$ takes the following simple form
\begin{center}
\begin{math}M_{IJ}= \bordermatrix{& & \cr
                                  &-I_{p+1}&0 \cr      
                                  &0  & I_{7-p}\cr}.   
\end{math}
\end{center}
The corresponding matrix in the fermionic sector is $M_{ab}=(\g^{1}\g^{2}\cdots \g^{p+1})_{ab}$. In the type IIA superstring the fermionic matrix $M_{a\dot{b}}$ contains an odd number of $\g$'s. These matrices express the rotation suffered by the superstring fields on the boundary.

In order to determine the solution to the equations (\ref{zeroTbc}) one can write them in terms of the oscillation modes of the superstring. The bosonic boundary conditions can be written as
\bea
\hat{p}^{I}|B\rangle = \left( \hat{x}^I - x^I \right)|B\rangle = 0~~,~~ 
\left( \a^I_n - M^{I}_{J}\bar{\a}^{J}_{-n}\right)|B\rangle = 0, \label{zeroTbbc} 
\eea
where $\hat{x}^I$ and $\hat{p}^I$ are the components of the position and momentum operators of the center of mass of the superstring. The ansatz for the fermionic oscillation modes on the boundary is
\be
\left(S^{a}_n + iM_{ab}\bar{S}^{b}_{n}\right)|B\rangle = 0~~,~~
\left(S^{\dot{a}}_n + iM_{\dot{a}\dot{b}}\bar{S}^{\dot{b}}\right)|B\rangle =0.\label{zeroTfbc}
\ee
These equations represent the boundary conditions in the physical Fock space of the superstring. The type IIB BPS $D$-brane state that solves them has the form of a coherent state \cite{gg}
\bea
|B\rangle &=& 
\exp \sum_{n>0}\left( M_{IJ}a^{I\dagger}_{n}\bar{a}^{J\dagger}_{n} - iM_{ab}S^{a\dagger}_{n}\bar{S}^{b\dagger}_{n}     \right)|B_0\rangle,\label{zeroTDbrane}\\
|B_0\rangle &=& (M_{IJ}|I\rangle|J\rangle +iM_{\dot{a}\dot{b}}|\dot{a}\rangle |\dot{b}\rangle)\label{zeroTDbranezero}.
\eea
Here, we have switched to the normalized form of the bosonic oscillator operators
\be
a^{I}_{n} = \frac{1}{\sqrt{n}}\a^{I}_n~,~a^{I\dagger}_{n}=\frac{1}{\sqrt{n}}\a^{I}_{-n}~,~n>0.
\label{bososc}
\ee
The zero mode factor defined in (\ref{zeroTDbranezero}) is obtained from the degenerate superstring vacuum and it is annihilated by all annihilation operators from the bosonic as well as fermionic sectors. In the absence of the open superstring condensates it can be normalized to unity \cite{gg}. However, the BPS $D$-brane does not belong to the physical Hilbert space of the superstring because it has an infinite norm.

\section{Thermal $D$-Brane Boundary States}

In order to determine the thermal boundary states from the BPS $D$-branes we are going to apply the TFD formalism which is better suited to study the structure of the thermal states. Also, the symmetry and supersymmetry breaking is manifest in the TFD approach \cite{ubook}. In the first subsection we are going to discuss the total system made of superstring degrees of freedom and the relevant degrees of freedom of the thermal reservoir. After that, we will derive the boundary conditions at finite temperature from the BPS $D$-brane boundary conditions at zero temperature. Next, we will show that a solution can be find by thermalizing the boundary states at $T=0$. 

\subsection{The Total System at $T=0$}

When the GS superstring is put in contact with a thermal reservoir, its physical properties are described by a $2d$ field theory at finite temperature. One way to understand the string heating is to imagine that each of its oscillators is heated separately by a corresponding oscillator degree of freedom of the reservoir. The situation is simpler to deal with in the light-cone gauge since the oscillators are independent. This picture is particularly useful if one wants to understand how the $D$-branes do heat since in the perturbative limit of string theory the $D$-branes are represented by the boundary states (\ref{zeroTDbrane}) as seen in the previous section. 

In the TFD approach \cite{ubook}, the degrees of freedom of the thermal reservoir can be described by an identical copy of the system and the putting the reservoir in contact with the system is called {\em doubling the system} (see the Appendix A for a quick glance at the TFD method.) The quantities related to the reservoir will be denoted by a $\ \tilde{~} \ $. 
The physical states of the thermal string belong to the thermal Hilbert space which is obtained from the total Hilbert space of the superstring and the tilde-superstring by a particular temperature dependent transformation.
Since one cannot tell which of two is actually the physical system and which stands for the degrees of freedom of the reservoir, $\ \tilde{} \ $ should be seen as denoting rather a symmetry of the total system. 
The tilde-conjugation is realized as an operation on the states and as an involution on the algebra of the operators of the doubled system defined by the axioms (\ref{tilde1})-(\ref{tilde4}) from Appendix A. The doubled system corresponding to the type II superstring in the light-cone gauge has the following (total) action
\be
\hat{\SS}_{l.c.g.} = \SS_{l.c.g.} - \tilde{\SS}_{l.c.g.},
\label{totalaction}
\ee
where $\SS_{l.c.g.}$ was given in the equation (\ref{stringaction}). Care should be taken when constructing the action $\tilde{\SS}_{l.c.g.}$ since it should be consistent with the tilde-conjugation axioms. By choosing the representation of the Dirac matrices as in the previous section one can show that $\tilde{\SS}_{l.c.g.}$ has the same form as $\SS_{l.c.g.}$ with all fields replace by the tilde-fields. The action (\ref{totalaction}) describes the superstring and the degrees of freedom of the thermal reservoir at zero temperature. Apart from the minus sign, the two systems are treated as independent (no interaction term is needed). In particular, the tilde-fields equations of motion derived from the above action are the Klein-Gordon equation and the Dirac equation. Also, we may conclude that the supersymmetry transformations of the tilde-superstring should be the same as those of the superstring, i. e. 
\bea
\delta \tilde{X}^{I} &=& \sqrt{2/p^{+}}\left( \g^{I}_{a\dot{a}}\tilde{\epsilon}^{*\dot{a}}\tilde{S}^{a} +
\g^{I}_{a\dot{a}}\bar{\tilde{\epsilon}}^{*\dot{a}}\bar{\tilde{S}^{a}}\right), \label{susytildeX}\\
\delta \tilde{S}^{a} &=& \sqrt{2 p^{+}}\left( \tilde{\eta}^{a} - i \rho^{\a}\6_{\a}
\tilde{{\cal X}}^{I}\g^{I}_{a\dot{a}}\tilde{\epsilon}^{\dot{a}}\right),
\label{susytildeS}\\
\delta \tilde{\bar{S}}^{a} &=& \sqrt{2 p^{+}}\left( \tilde{\bar{\eta}}^{a} - i \rho^{\a}\6_{\a}
\tilde{\bar{\cal X}}^{I}\g^{I}_{a\dot{a}}\tilde{\bar{\epsilon}}^{\dot{a}}\right).
\label{susybartildeS}
\eea 

The ground state $|\phi_0\rangle$ of the type IIB superstring at $T = 0$ contains four sectors which are obtained from the tensor product of the left-moving and right-moving bosonic and fermionic massless states
\be
|\phi_0\rangle = \{ |I\rangle|J\rangle,\, |I\rangle|b\rangle,\, |a\rangle|J\rangle,\, |a\rangle|b\rangle \}.
\label{zeroTgroundstate}
\ee
The states $|I\rangle$ and $|a\rangle$ belong to $\bold{8}_v$ and $\bold{8}_s$ representations of $SO(8)$. They are obtained from the oscillator vacuum $|0\rangle_l$ in the left-moving sector as follows
\be
|I\rangle = a^{I\dagger}_{1}|0\rangle_l~,~|a\rangle = \Theta^{a}|0\rangle_l,
\label{groundstate1}
\ee
where the spin operator satisfies $\Theta^{a} = \lim_{z\rightarrow 0}\Theta^{a}(z)$. The ground state in the right-moving sector is obtained in a similar fashion. (In the type IIA superstring spinors of different chirality should be considered.) The ground state of the tilde-superstring $|\tilde{\phi}_0\rangle$ is defined by the same relations with the fields replaced by tilde-fields. The ground state of the total system 
$|\phi_0\rangle \rangle$ should be tilde-invariant. However, if we take the tensor product $|\phi_0\rangle|\tilde{\phi}_0\rangle$ the products among different sectors are tilde-invariant only if the order of factors is irrelevant, e. g. the states $|I\rangle|J\rangle|\tilde{a}\rangle|\tilde{J}\rangle$ and $|\tilde{I}\rangle|\tilde{J}\rangle|a\rangle|J\rangle$ are mapped into one another only if one can permute freely the states from tilde and non-tilde Hilbert spaces. This shows that there are two possibilities to extend the tilde operation to the vector states: i) by applying the axioms (\ref{tilde1})-(\ref{tilde4}) to the components and taking simultaneously the vector transpose or ii) by applying the axioms without transposing the column vector. However, if one requires that each sector of the total ground state be tilde-invariant, then only the diagonal sectors from the product $|\phi_0\rangle|\tilde{\phi}_0\rangle$ should be taken into account. This condition can be written as 
\bea
|\phi_{0}\rangle\rangle &=& \mbox{diag}(|\phi_0\rangle|\tilde{\phi}_0\rangle ), \label{totalgs1}\\
\left( |\phi_{0}\rangle\rangle \right)\tilde{} &=& |\phi_{0}\rangle\rangle. \label{totalgs2} 
\eea
From this we see that the sectors of the total thermal ground state are 
\be
|\phi_{0}\rangle\rangle = \{ |I\rangle|J\rangle|\tilde{I}\rangle|\tilde{J}\rangle ,
|a\rangle|J\rangle|\tilde{a}\rangle|\tilde{J}\rangle ,
|I\rangle|b\rangle|\tilde{I}\rangle|\tilde{b}\rangle ,
|a\rangle|b\rangle|\tilde{a}\rangle|\tilde{b}\rangle \} \equiv \{ |\phi\rangle\rangle | \bar{\phi}\rangle\rangle \}.
\label{totalgs}
\ee 
The state (\ref{totalgs}) is tilde-invariant and consistent with the Kronecker product \footnote{Note that due to the symmetry of the states in each sector, the Kronecker product does not impose any new restriction on the content of sectors.}. In the second equality we have used the notations $ | \phi \rangle\rangle = | \phi \rangle |\tilde{\phi} \rangle $ and 
$|\bar{\phi}\rangle\rangle = |\bar{\phi}\rangle |\tilde{\bar{\phi}} \rangle$, where $\phi = \{ I, a \} $. This is basically a rearrangement of terms, e. g. 
\be
|I\rangle|J\rangle|\tilde{I}\rangle|\tilde{J}\rangle = (|I\rangle|\tilde{I}\rangle)(|J\rangle|\tilde{J}\rangle )
\equiv |I\rangle\rangle|J\rangle\rangle. \label{compactgs}
\ee
The most natural way to impose BPS $D$-brane boundary conditions to the total superstring is by requiring that the relations (\ref{zeroTbc}) and a similar set in which the superstring fields are replaced by the tilde-superstring fields hold for the total theory. Heuristically, this choice is motivated by the fact that the hypersurface that defines the $D$-brane continues to be present in spacetime at the same location after the superstring is put in contact with the thermal reservoir. Implementing the boundary conditions in the total Fock space is straightforward except for a subtlety concerning the ansatz corresponding to the relations (\ref{zeroTbbc}) and (\ref{zeroTfbc}). To see that, note that due to the nontrivial action of the tilde-conjugation on the complex numbers, the wave functions corresponding to the negative and positive modes in the Fourier expansion of the bosonic and fermionic solutions (\ref{Fourierbos}) and (\ref{leftrightFourierferm}) are interchanged in the expansion of the corresponding tilde-fields. This result can be written in a compact form by introducing the thermal doublet fields \cite{ubook} 
\bea
\hat{X}^{I}(\t ,\s ) &=& \hat{x}^{I} + 2\a '\hat{p}^{I}\tau + i\sqrt{\frac{\a '}{2}}\sum_{n \neq 0}\frac{1}{n} \left( \hat{\a}^{I}_n e^{-2in(\t -\s )} + \hat{\bar{\a}}^{I}_n e^{-2in(\t + \s )} \right),\label{hatFourierboszero}\\
\hat{S}^{1a}(\t ,\s ) &=& \sum_{n}\hat{S}^{a}_n e^{-2in(\t -\s )}~~,~~
\hat{S}^{2a}(\t ,\s ) = \sum_{n}\hat{\bar{S}}^{a}_n e^{-2in(\t +\s )},\label{hatFourierfermzero}
\eea
where the thermal doublet oscillators are defined as
\bc
\begin{math}
\hat{\a}^{I}_{n} = \bordermatrix{& \cr          
                      & \a^{I}_n  \cr          
                      & \tilde{\a}^{I}_{-n}  \cr}
~~
\hat{\a}^{I}_{-n} = \bordermatrix{& \cr          
                      & \a^{I}_{-n}  \cr          
                      & \tilde{\a}^{I}_{n}  \cr}
~~
\hat{\bar{\a}}^{I}_{n} = \bordermatrix{& \cr          
                      & \bar{\a}^{I}_n  \cr          
                      & \tilde{\bar{\a}}^{I}_{-n}  \cr}
~~
\hat{\bar{\a}}^{I}_{-n} = \bordermatrix{& \cr          
                      & \bar{\a}^{I}_{-n}  \cr          
                      & \tilde{\bar{\a}}^{I}_{n}  \cr},
\label{thermaldoublebososc}
\end{math}
\ec
for all $n > 0$ in the bosonic sector and 
\bc
\begin{math}
\hat{S}^{a}_{n} = \bordermatrix{& \cr          
                      & S^{a}_n  \cr          
                      & \tilde{S}^{a}_{-n}  \cr}
~~,~~
\hat{S}^{a}_{-n} = \bordermatrix{& \cr          
                      & S^{a}_{-n}  \cr          
                      & \tilde{S}^{a}_{n}  \cr}
~~,~~
\hat{\bar{S}}^{a}_{n} = \bordermatrix{& \cr          
                      & \bar{S}^{a}_n  \cr          
                      & \tilde{\bar{S}}^{a}_{-n}  \cr}
~~,~~
\hat{\bar{S}}^{a}_{-n} = \bordermatrix{& \cr          
                      & \bar{S}^{a}_{-n}  \cr          
                      & \tilde{\bar{S}}^{a}_{n}  \cr},
\label{thermaldoublefermosc}
\end{math}
\ec
for $ n \geq 0$ in the fermionic sector. In terms of the thermal doublets the boundary conditions for the total superstring can be written in the folowing compact form
\bea
\left(\hat{\a}^{I}_{n} - \hat{M}^{I}_{J}\hat{\bar{\a}}^{J}_{-n}\right)|B\rangle\rangle &=& 0,\label{firsthatbosbc}\\
\left(\hat{\a}^{I}_{-n} - \hat{M}^{I}_{J}\hat{\bar{\a}}^{J}_{n}\right)|B\rangle\rangle &=& 0,\label{secondhatbosbc}
\eea
for all $n > 0$ in the bosonic sector. The ansatz in the fermionic sector of the superstring naturally extends to the fermionc sectors of the total superstring to  
\bea
\left(\hat{S}^{a}_{n} + i \hat{M}_{ab}\hat{\bar{S}}^{b}_{-n}\right)|B\rangle\rangle &=& 0,\label{firsthatfermbc}\\
\left(\hat{S}^{a}_{-n} + i \hat{M}_{ab}\hat{\bar{S}}^{b}_{n}\right)|B\rangle\rangle &=& 0,\label{secondhatfermbc}
\eea
for all $n > 0 $ in the fermionic sector. The matrices $ \hat{M}^{I}_{J}$ and $ \hat{M}^{a}_{b} $ have the following form
\begin{center}
\begin{math}
\hat{M}^{I}_{J}=
\bordermatrix{& & \cr
              &M^{I}_{J}&0 \cr      
              &0  & M^{I}_{J}\cr}
~~,~~                 
\hat{M}_{ab}= 
\bordermatrix{& & \cr
              &M_{ab}&0 \cr      
              &0  & - M_{ab}\cr}.  
\label{hatmatrices}
\end{math}
\end{center}
The difference between the boundary conditions for the superstring sector and the tilde-superstring consists in the change of $n$ and $-n$ indices and in a relative minus sign in the $M$ matrix in the spinor sector. Nevertheless, since $n$ runs over all integers less zero, the boundary conditions in the two systems will eventually coincide. This suggests that the vector $| B \rangle\rangle $ from the total Fock space be factorized as
\be
| B \rangle\rangle = |B\rangle \otimes \widetilde{|B\rangle} \equiv |B\rangle \widetilde{|B\rangle} .
\label{factortotalbrane}
\ee 
The explicit form of the $D$-brane state of the total system can be obtained either from the direct calculations or from the tilde-conjugation of $|B\rangle$ state and it is given by the following relation
\be
|B\rangle\rangle = 
\exp \sum_{n>0}\left[ M_{IJ}(a^{I\dagger}_{n}\bar{a}^{J\dagger}_{n} + \tilde{a}^{I\dagger}_{n}\tilde{\bar{a}}^{J\dagger}_{n})
- iM_{ab}(S^{a\dagger}_{n}\bar{S}^{b\dagger}_{n} - \tilde{S}^{a\dagger}_{n}\tilde{\bar{S}}^{b\dagger}_{n}) \right]|B_0\rangle\rangle ,
\label{totalDbrane}
\ee
where the total brane ground state is the tensor product
\be
|B_0\rangle\rangle = (M_{IJ}|I\rangle|J\rangle + iM_{\dot{a}\dot{b}}|\dot{a}\rangle |\dot{b}\rangle)
\otimes
(\tilde{M}_{IJ}\widetilde{|I\rangle}\widetilde{|J\rangle} - i\tilde{M}_{\dot{a}\dot{b}}\widetilde{|\dot{a}\rangle}
\widetilde{|\dot{b}\rangle}).
\label{totalzeroDbrane}
\ee
The minus sign in the tilde-state is due to the tilde conjugation. Alternatively, one can show that for the tilde-superstring, the conservation of half of the supercharges is obtained by applying $\tilde{Q}^{-a}$ and $\tilde{\bar{Q}}{}^{-\dot{a}}$ operators on the Fock space. The total $D$-brane state given above represents a condensate of the type II superstring oscillators and the thermal reservoir degrees of freedom localized on the hypersurface defined by the Dirichlet and the Neumann boundary conditions imposed on both superstring and reservoir.

\subsection{Thermalization of the Superstring}

According to the TFD, the superstring will be heat up to the temperature $T \neq 0$ by a particular interaction between all pairs of identical modes from the non-tilde and tilde sectors of the total superstring. This interaction is represented by the temperature dependent Bogoliubov operator that acts on the total Hilbert space and it is called {\em thermalization}. The Bogoliubov operator is constructed out of pairs of creation and annihilation operators of identical oscillators from non-tilde and tilde-sectors and it maps the total Hilbert space $\hat{\HH}$ to the thermal Hilbert space $\HH(\b_T)$. In general, for systems with an infinite number of degrees of freedom the two spaces are not isomorphic \cite{ubook}. 

The Hilbert space of the total superstring has the structure of a tensor product
\be
\hat{\HH} = \HH_{\small c.m.} \otimes (\HH_0 \otimes \tilde{\HH}_0) \otimes (\otimes_n {\HH}^{b}_{n})\otimes (\otimes_n\tilde{\HH}^{b}_n) \otimes
(\otimes_n {\HH}^{f}_{n})\otimes (\otimes_n\tilde{\HH}^{f}_n),
\label{TotalHilbertTensor}
\ee
where $\HH_{\small c.m.}$ is the Hilbert space of the center of mass, $\HH_0$ and $\tilde{\HH}_0$ are the Hilbert spaces of the fermionic zero modes and $\HH^{b,f}_n$ and $\tilde{\HH}^{b,f}_n$ are the oscillators in the superstring and tilde-superstring sectors, respectively. Here, the superscripts $b$ and $f$ stand for fermions and bosons, respectively. 

The Bogoliubov operator for the type II superstring has a simple form in the light-cone gauge. Indeed, since the oscillators represent physical degrees of freedom only, the Bogoliubov operator is just the algebraic sum of the Bogoliubov operators corresponding to all fermionic and bosonic oscillators and it acts on the Fock space of the total system. The construction of the temperature dependent Bogoliubov operator $G$ and its action on all superstring creation and annihilation operators are given in the Appendix B. From its definition (\ref{BogOp})
we see that it generates an unitary and tilde-invariant transformation. Let us denote the set of all oscillator operators by
\be
\OO = \{ O \} \equiv \{ a^{I}_n, a^{I \dagger}_n, \tilde{a}^{I}_n, \tilde{a}^{I\dagger}_n;\
\bar{a}^{I}_n, \bar{a}^{I \dagger}_n, \tilde{\bar{a}}^{I}_n, \tilde{\bar{a}}^{I\dagger}_n;\ 
S^{a}_n, S^{a \dagger}_n, \tilde{S}^{a}_n, \tilde{S}^{a\dagger}_n;\ 
\bar{S}^{a}_n, \bar{S}^{a \dagger}_n, \tilde{\bar{S}}^{a}_n, \tilde{\bar{S}}^{a\dagger}_n \}. 
\label{alloscillators}
\ee
By the similarity transformation generated by $G$ a new set of thermal oscillators is obtained 
\be
\OO (\beta_T) = e^{-iG}\, \OO \, e^{iG} \equiv \{ e^{-iG}\, O \, e^{iG} \}.
\label{Toscillators}
\ee
The thermalization does not affect the wave-functions. Thus, the operators (\ref{Toscillators}) can be viewed as the Fourier components of the superstring coordinates
\bea
X^{I}(\b_T ) = e^{-iG}\, X^{I} \, e^{iG}~ , ~ S^{a}(\b_T ) = e^{-iG}\, S^{a} \, e^{iG}~,~ \bar{S}^{a}(\b_T ) = e^{-iG}\, \bar{S}^{a} \, e^{iG}, 
\label{thermalstringops}\\
\tilde{X}^{I}(\b_T ) = e^{-iG}\, \tilde{X}^{I} \, e^{iG} ~,~ \tilde{S}^{a}(\b_T ) = e^{-iG}\, \tilde{S}^{a} \, e^{iG} ~,~ \tilde{\bar{S}}^{a}(\b_T ) = e^{-iG}\, \tilde{\bar{S}}^{a} \, e^{iG}. \label{thermaltildestringops}
\eea
That shows that the effect of the thermalization of the superstring oscillators is equivalent to the thermalization of the superstring fields. In particular, the thermal string coordinates satisfy the equations of motion derived from the following Lagrangian 
\be
\LL(\b_T ) = e^{-iG}\hat{\LL}e^{iG}.
\label{thermallag}
\ee

Let us turn now to the thermal ground state. According to the TFD method, the thermal vacuum state 
$|0(\b_T )\rangle\rangle$ is defined as
\be
|0(\b_T )\rangle\rangle = e^{-iG}|0\rangle\rangle~,~|0\rangle\rangle = |0\rangle|\tilde{0}\rangle,
\label{thermvacuum}
\ee
where $|0\rangle$ and $|\tilde{0}\rangle$ denote the vacuum state of the superstring and tilde-superstring, respectively. The above state is annihilated by all thermal bosonic and fermionic annihilation operators. We observe that one should be able to obtain the thermal state $| \phi_0(\b_T )\rangle\rangle$ from the thermal vacuum $|0(\b_T)\rangle\rangle$ by acting upon it with the operators  $a^{I\dagger}_{1}(\b_T )$, $\bar{a}^{I\dagger}_{1}(\b_T )$, $\Theta^{a}(\b_T)$ and $\bar{\Theta}^{a}(\b_T)$ and the corresponding tilde-operators in the same way as the superstring ground state is obtained from the superstring vacuum. 
The $a^{I\dagger}_{1}(\b_T )$ and $\bar{a}^{I\dagger}_{1}(\b_T )$ are obtained by applying the similarity transformation (\ref{Toscillators}) to the corresponding operators at zero temperature. However, there is no definite TFD prescription for the thermalization of the spin operators. Nevertheless, the most natural way to transform the ground state at $T \neq 0$ in the bosonic and the fermionic sectors is to consider the same transformation for the spin operators as for the oscillator operators. Thus, we make the assumption that the spin operators transform according to the following relations
\bea
\Theta^{a} (\b_T ) &=& e^{-iG}\Theta^{a}e^{iG}~,~\bar{\Theta}^{a} (\b_T ) = e^{-iG}\bar{\Theta}^{a}e^{iG},\label{thermTheta}\\
\tilde{\Theta}^{a} (\b_T ) &=& e^{-iG}\tilde{\Theta}^{a}e^{iG}~,~
\tilde{\bar{\Theta}}^{a} (\b_T ) = e^{-iG}\tilde{\bar{\Theta}}^{a}e^{iG}.
\label{thermatildeTheta}
\eea
The bosonic components of the thermal ground state can be written as
\bea
|I(\b_T)\rangle\rangle &=& a^{I\dagger}_{1}(\b_T)|0(\b_T)\rangle\rangle~~,~~
|\bar{I}(\b_T)\rangle\rangle = \bar{a}^{I\dagger}_{1}(\b_T)|0(\b_T)\rangle\rangle,\label{boscompground1}\\
|\tilde{I}(\b_T)\rangle\rangle &=& \tilde{a}^{I\dagger}_{1}(\b_T)|0(\b_T)\rangle\rangle~,~
|\tilde{\bar{I}}(\b_T)\rangle\rangle = \tilde{\bar{a}}^{I\dagger}_{1}(\b_T)|0(\b_T)\rangle\rangle.
\label{boscompground2}
\eea
For the fermionic components of the thermal ground state the following relations hold
\bea
|a(\b_T)\rangle\rangle &=&\Theta^{a}(\b_T) |0(\b_T)\rangle\rangle~~,~~
|\bar{a}(\b_T)\rangle\rangle =\bar{\Theta}^{a}(\b_T) |0(\b_T)\rangle\rangle,\label{fermcompground1}\\
|\tilde{a}(\b_T)\rangle\rangle &=&\tilde{\Theta}^{a}(\b_T) |0(\b_T)\rangle\rangle~~,~~
|\tilde{\bar{a}}(\b_T)\rangle\rangle =\tilde{\bar{\Theta}}^{a}(\b_T) |0(\b_T)\rangle\rangle.
\label{fermcompground2}
\eea
The thermal ground state is obtained by taking the tilde-invariant component of the tensor product of the above 
states. Using the compact form of the total ground state given in the relations (\ref{totalgs}) and (\ref{compactgs})
one can write the thermal ground state as
\be
|\phi_{0}(\b_T)\rangle\rangle = e^{-iG}|\phi_0\rangle\rangle \equiv e^{-iG}\{ |\phi\rangle\rangle
|\bar{\phi}\rangle\rangle \} 
\label{compactgsT}.
\ee
Note that, by construction, the thermal ground state contains only massless excitations of the thermal fields. However, in terms of fields at zero temperature, it contains a mixture of massless and massive modes. 

\subsection{Thermal Boundary States}

We have seen that the thermalization of the superstring is represented by an unitary transformation acting on the total Hilbert space and an induced similarity transformations acting on the oscillator operators. Therefore, the superstring zero temperature boundary conditions (\ref{zeroTbc}) and their corresponding tilde-superstring boundary conditions are mapped by the thermalization procedure to the following set of equations
\bea
(\partial X^{I}(\b_T)-M_{J}^{I}\overline{\partial}X^{J}(\b_T))|B(\b_T)\rangle\rangle=0,\label{Tbctotalstring1}\\
Q^{+a}(\b_T)|B(\b_T)\rangle \rangle = Q^{+\dot{a}}(\b_T)|B(\b_T)\rangle\rangle=0\label{Tbctotalstring2},\\
(\partial \tilde{X}^{I}(\b_T)-M_{J}^{I}\overline{\partial}\tilde{X}^{J}(\b_T))|B(\b_T)\rangle\rangle=0,
\label{Tbctotaltildestring1}\\
\tilde{Q}^{-a}(\b_T)|B(\b_T)\rangle \rangle = \tilde{Q}^{-\dot{a}}(\b_T)|B(\b_T)\rangle\rangle=0 .\label{Tbctotaltildestring2}
\eea
We consider the above equations as representing the thermal boundary conditions of the thermal string\footnote {Alternatively, they can be obtained from the thermal lagrangian given in the relation (\ref{thermallag}).}. Their solution denoted by $|B(\b_T )\rangle\rangle$ represents the thermal boundary state and belongs to the thermal Hilbert space $\HH(\b_T)$. To obtain its explicit form note that if $|B\rangle\rangle$ is a solution of (\ref{zeroTbc}) and the corresponding tilde boundary conditions then the following vector 
\be
|B(\b_T)\rangle\rangle = e^{-iG}|B\rangle\rangle
\label{thermalsolution}
\ee
represents a solution of the equations (\ref{Tbctotalstring1}), (\ref{Tbctotalstring2}), (\ref{Tbctotaltildestring1}) and (\ref{Tbctotaltildestring2}). This shows that the thermal boundary state is obtained from the $D$-brane state of the total system by thermalization. Due to the independence of the superstring from the tilde-superstring, $|B\rangle\rangle$ was written as a tensor product
in the relation (\ref{factortotalbrane}). Therefore, the general form of the thermalized boundary state is given by the following relation
\be  
|B(\b_T)\rangle\rangle = e^{\Sigma(\b_T) + \tilde{\Sigma}(\b_T)}|B_0(\b_T)\rangle\rangle,
\label{thermalsol1}
\ee
where we have introduced the following notations
\bea
\Sigma(\b_T) &=& \sum_{n>0}( M_{IJ}a^{I\dagger}_{n}(\b_T)\bar{a}^{J\dagger}_{n}(\b_T) - iM_{ab}S^{a\dagger}_{n}(\b_T)\bar{S}^{b\dagger}_{n}(\b_T)), \label{sigmaop}\\    
\tilde{\Sigma}(\b_T) &=& \sum_{n>0}( \tilde{M}_{IJ}\tilde{a}^{I\dagger}_{n}(\b_T)\tilde{\bar{a}}^{J\dagger}_{n}(\b_T) + i\tilde{M}_{ab}\tilde{S}^{a\dagger}_{n}(\b_T)\tilde{\bar{S}}^{b\dagger}_{n}(\b_T)),\label{tildesigmaop}\\
|B_0(\b_T)\rangle\rangle &=& e^{-iG}|B_0\rangle\rangle.\label{thermalzeromode1}
\eea
The thermal boundary state (\ref{thermalsol1}) is tilde-invariant and it is expressed in terms of components of the degenerate ground state of the total superstring at zero temperature. The natural question is whether the thermal boundary state can be generated from the thermal ground state. To this end, one has to express $|B_0(\b_T)\rangle\rangle$ in terms of components of the thermal boundary state instead of $|B_0\rangle\rangle$. Using the relations from Appendix B one can easily see that
\be
|B_0(\b_T)\rangle\rangle = (M_{IJ}|IJ(\b_T)\rangle\rangle +iM_{\dot{a}\dot{b}}|\dot{a}\dot{b}(\b_T)\rangle\rangle)
(M_{IJ}|\tilde{I}\tilde{J}(\b_T)\rangle\rangle -iM_{\dot{a}\dot{b}}|\tilde{\dot{a}}\tilde{\dot{b}}(\b_T)\rangle\rangle).
\label{thermalzeromode3}
\ee
Here, $|IJ(\b_T)\rangle\rangle$ denote the state obtained by acting with $a^{I\dagger}_{1}(\b_T)\bar{a}^{J\dagger}_{1}(\b_T)$ on the ground state $|0(\b_T)\rangle\rangle$.

\section{Properties of the Thermal String}

In this section we are going to determine the supersymmetry content of the thermal string. We will verify that the supersymmetry is broken in the thermal vacuum state and in the thermal ground state. The same result extends to the thermal boundary states. Next, we will compute the entropy of the thermal string in the TFD approach.

\subsection{Supersymmetry Breaking in the Thermal String}

According to the TFD, the symmetry content of the thermal string theory is checked by applying the symmetry generators at $T = 0$ from the non-tilde sector on to the thermal states \cite{ubook}. Let us consider the right-moving supersymmetry charges in terms of the Fourier components given in the relations (\ref{susygenQ}) and (\ref{susygenbarQ}). The $\eta$-transformation is generated only by the zero fermionic mode which implies that   
\be
\langle\langle 0(\b_T )| Q^{a}_{\eta}|0(\b_T )\rangle\rangle = \langle\langle 0 | Q^{a}_{\eta}|0 \rangle\rangle.
\label{etathermvac}
\ee    
The $\epsilon$-transformation is generated by a mixture of nonzero bosonic and fermionic modes. It is easy to check that the following commutation relations hold
\be
[ a^{I\dagger}_{n}, G ] = i\theta^{B}_{n}\tilde{a}^{I}_{n}~,~[S^{a\dagger}_{n}, G ] = i\theta^{F}_{n}\tilde{S}^{a}_{n} .
\label{bosfermcomm} 
\ee
This shows that the bosonic and fermionic oscillator operators are not invariant under thermalization. Therefore, the $\epsilon$-supersymmetry is broken 
\be
\langle\langle 0(\b_T )| Q^{a}_{\epsilon}|0(\b_T )\rangle\rangle \neq \langle\langle 0 | Q^{a}_{\epsilon}|0 \rangle\rangle.
\label{epsilonthermvac}
\ee    
One can apply the same reasoning to the thermal ground state, too. By using the form of  
$|\phi_{0}(\b_T)\rangle\rangle$ given in the equation (\ref{compactgsT}) one can easily check that 
\be
\langle\langle \phi_{0}(\b_T) | Q^{a}_{\eta}| \phi_{0}(\b_T)\rangle\rangle =  \langle\langle \phi_{0} | Q^{a}_{\eta}|\phi_{0}\rangle\rangle~,~
\langle\langle \phi_{0}(\b_T) | Q^{a}_{\epsilon}| \phi_{0}(\b_T)\rangle\rangle \neq \langle\langle \phi_{0} | Q^{a}_{\epsilon}|\phi_{0}\rangle\rangle.
\label{thermgroundbreak}
\ee    

The above pattern of supersymmetry breaking extends to the full thermal Hilbert space. Since, by assumption, the fermionic zero mode does not transform under the Bogoliubov transformation and it commutes with all other bosonic and fermionic modes, the $\eta$-supersymmetry is preserved. On the other hand, due to the non-trivial action of the Bogoliubov transformation on the non-zero superstring modes, the $\epsilon$-supersymmetry is broken. This entails the breaking of the full supersymmetry algebra in the thermal Hilbert space in each of the tilde and non-tilde sectors, separately. In particular, since $|B(\b_T)\rangle\rangle$ was obtained by thermalization according to the relation (\ref{thermalsolution}), the $\epsilon$-supersymmetry will be broken and the $\eta$-supersymmetry will be preserved in 
the thermal boundary state. 

\subsection{Thermal Partition Function and Entropy of the Thermal String}

The entropy of the thermal string is given by the expectation value of the entropy operator $K$ in the thermal vacuum. However, as argued in \cite{ivv1} and discussed in more detail in \cite{ng1,hemi}, the ansatz that defines the thermal vacuum $|0(\b_T )\rangle\rangle$ must be modified in order to take into account the constraints. In particular, since the constraints are already solved in the light-cone gauge, one has to implement the level matching condition. The main consequence of this modification appears in the calculation of the partition function $Z(\b_T )$.

The level matching condition for the GS superstring is given by the following relation
\be
L^{l.c.g.}_{0}=\bar{L}^{l.c.g.}_{0},
\label{lmc}
\ee
where the Virasoro operators have the following explicit form \cite{gsw}
\bea
L^{l.c.g.}_{0} &=& \frac{\a '}{4}p^{2} + \sum_{n=1}^{\infty} n a^{I\dagger}_{n} a^{I}_{n} +
\sum_{m>0}m S^{a\dagger}_{m} S^{a}_{m},\label{vir}\\
\bar{L}^{l.c.g.}_{0} &=& \frac{\a '}{4}p^{2} + \sum_{n=1}^{\infty} n \bar{a}^{I\dagger}_{n} \bar{a}^{I}_{n} + \sum_{m>0}m \bar{S}^{a\dagger}_{m} \bar{S}^{a}_{m}\label{barvir}.
\eea
As in the RNS formulation, the level matching condition implements the invariance of the theory under the world-sheet reparametrization under $\sigma \rightarrow \sigma + \e$. The partition function of the thermal string in the TFD formulation is given by the following relation
\be
Z(\b_T ) = \mbox{Tr} \left[ \delta(P = 0) e^{-\b_T H }\right],
\label{partfunct}
\ee 
where $P=L^{l.c.g.}_{0} - \bar{L}^{l.c.g.}_{0}$ is the world-sheet momentum and $H = L^{l.c.g.}_{0}+\bar{L}^{l.c.g.}_{0}$ is the Hamiltonian and the trace is taken over the full Hilbert space of the superstring. The delta function projects the trace on the physical subspace. One obtains after some algebra the following partition function for the thermal string
\be
Z(\b_ T) = \int_{-1/2}^{+1/2}ds 
\left[ \prod_{n=1}^{\infty} e^{2(\l(s,\b_T ) + \bar{\l}(s,\b_T ))n}
\frac{\left(1+e^{\l (s,\b_T)n} \right)\left(1+e^{\bar{\l} (s,\b_T)n} \right)}{\left(1-e^{\l (s,\b_T)n} \right)\left(1-e^{\bar{\l} (s,\b_T)n} \right)}
\right]^{8},
\label{partfunctfin}
\ee
where we have used the notations
\be
\l (s,\b_T) = 2 \pi i s - \frac{\b_T}{2p^+}~~,~~
\bar{\l} (s,\b_T) = 2 \pi i s + \frac{\b_T}{2p^+}.
\label{lambdas}
\ee
The relation (\ref{partfunctfin}) contains contributions from both bosonic and fermionic sectors.
Also, we have taken $\o^{B}_{n} = n$ and $\o^{F}_{n}=n$ for the frequencies of the bosonic and fermionic oscillators in $\hbar$ units. Note that the relation (\ref{partfunctfin}) defines the transverse partition function at fixed $p^{+}$. The full partition function is obtained from $Z (\b_T)$ as \cite{eo}
\be
\ZZ (\b_T ) = \frac{L}{\sqrt{2 \pi }} \int dp^{+} e^{\b_T p^{+}} Z(\b_T, p^{+}),
\label{fullpf}
\ee
where $L$ is the length in the longitudinal direction $Z(\b_T, p^{+})$ is $Z(\b_T)$ with the dependence on $p^{+}$ restored.

By using the formulas from the Appendix A and after expressing the operators at $T \neq 0$ in terms of operators at $T = 0$, one can show that the bosonic and fermionic contributions to the entropy, in $k_B$-units can be written as
\be
S^{B} = -16 \sum^{\infty}_{n=1} \log \{ \xi_{n} \left( 1-\xi_{n}\right)^{\f{1}{\xi_n}} \}~,~ 
S^{F} = -16 \sum^{\infty}_{n=1} \log \{ \chi_{n} \left( \chi_{n}-1 \right)^{\f{\chi_n-1}{\chi_n}} \},\label{bosfermentr}
\ee
where 
\be
\xi_{n} = 1 - e^{-\b_T n}~,~\chi_{n} = 1+ e^{\b_T n}.
\label{xichi}
\ee
The bosonic and fermionc entropies given in the relation (\ref{bosfermentr}) show that the entropy of the thermal string goes to zero as $T \rightarrow \infty$. This shows that the perturbative expansion of superstring breaks down at higher temperatures, which agrees with the current understanding of the superstring at finite temperature \cite{aw}.

\section{Conclusions and Discussions}

In this paper we have shown that the boundary state formalism used to obtain the type II $D$-branes at zero temperature can be naturally extended to finite temperature in the light-cone gauge and in the TFD approach. In this framework, the superstring is thermalized to the thermal string and the thermal $D$-branes are obtained by thermalization of the $D$-branes at zero temperature. During the heating of the superstring, the supersymmetry is broken as expected. In particular, the $\eta$-transformation is preserved at finite temperature while the $\epsilon$-transformation is not. This supersymmetry breaking pattern is repeated for all states from the thermal Hilbert space, including the thermal $D$-branes.  Nevertheless, since the information on the way in which the supersymmetry is broken is now available, it would be interesting to see whether thermal boundary states that preserve the $\epsilon$-supersymmetry can be constructed. (Note that quantities of the form $O-\tilde{O}$ are preserved by the Bogoliubov transformations.) Also, we have computed the partition function and the entropy of the thermal string in the thermal vacuum $|0(\b_T )\rangle\rangle$ according to the TFD prescription. This state is a mixture of massless and massive zero temperature superstring states. The entropy tends to zero for higher temperatures (see also \cite{ng1} for a derivation of the string thermodynamics in $pp$-wave background) which demonstrates a behavior consistent with the interpretation that the perturbative expansion of the superstring theory does not work at high temperature. If one computes the entropy of the bosonic and fermionic superstring oscillations in the thermal $D$-brane state one can show that it diverges for all $T$'s. However, this is a general feature of TFD which, in general, leads to infinite entropies. More important is the question whether such calculations have a sensible interpretation since, in order to define the $D$-brane entropy in a TFD fashion, one should be able to develop firstly a canonical formalism for the $D$-branes. 

The method presented in this paper can be applied to study the thermal string in other backgrounds, too, possibly by employing the path integral formulation of the TFD formalism. However, in order to study the thermal $D$-branes in the boundary state formalism, the canonical quantization should be available. As an interesting and non-trivial application, the calculation of the thermodynamical functions of the thermal bosonic string in the $AdS$ black-hole background in the first order perturbation canonical quantization is discussed in \cite{hemi}. The general canonical quantization of the string in the $AdS$ black-hole background was given in \cite{ns1,ns2} and it is based on the expansion of the string fields around the geodesic solution without expanding the metric. It is interesting to see whether this system has consistent boundary states with the low energy branes and whether they can be mapped at finite temperature. We hope to report on this problems in future.

{\bf Acknowledgments}

I would like to acknowledge to J. A. Helay\"{e}l-Neto, S. Alves and A. M. O. de Almeida for hospitality at 
LAFEX-CBPF where part of this work was done.

\appendix

\section{Review of the TFD Formalism}

The framework of the TFD consists in the following assumptions. Firstly, one has to duplicate the operators corresponding to the degrees of freedom of the system under consideration. Secondly, the properties of the thermal ground state must be specified (the thermal state condition.) 

The operator freedom is doubled by associating to any operator $A$ an operator $\tilde{A}$. The two sets $\{ A \}$ and $\{ \tilde{A} \}$ are considered independent
\be
[A,\tilde{B}]_{\pm}=0,
\label{indeptildeops}
\ee
for any $A$ from $\{A\}$ and any $\tilde{B}$ from $\{\tilde{A}\}$. Here, $+$ denotes the anti-commutator taken when both $A$ and $\tilde{B}$ are fermionic. Otherwise, the commutator should be taken. The two sets of operators $A$ and $\tilde{A}$ are in one-to-one correspondence through the tilde-conjugation rule given by the following axioms
\bea
(A_1 A_2 )\ \tilde{} &=& \tilde{A}_1\tilde{A}_2,\label{tilde1}\\
(c_1 A_1 + c_2 A_2 )\ \tilde{} &=& c^{*}_1 \tilde{A}_1 + c^{*}_2 \tilde{A}_2, \label{tilde2}\\ 
(A^{\dagger})\ \tilde{} &=& (\tilde{A})^{\dagger},\label{tilde3}\\
(\tilde{A})\ \tilde{} &=& A,\label{tilde4}
\eea
for all complex numbers $c$, $c_i$ and all operators $A$, $A_i$. Here, $c^*$ denotes the complex conjugate of the number $c$. The tilde-conjugation is an involution on the algebra of the total string operators.

The thermal ground state is specified by the thermal state condition. In general, this is defined by taking asymmetric bra and ket ground states that satisfy the following relations
\bea
\langle 1 | &=& \sigma^{*}\langle 1 | \tilde{A}^{\dagger},\label{thermalbra}\\
A|0(\b_T)\rangle &=& \sigma e^{\b_T \hat{H}}\tilde{A}^{\dagger}e^{-\b_T \hat{H}}|0(\b_T) \rangle,\label{thermalket}
\eea
for any $A$, where $\sigma = +1$ for bosons and $\sigma = i$ for fermions and 
\be
\hat{H} = H = \tilde{H}.
\label{totalHamiltonian}
\ee
The phases of the thermal ground states (thermal vacua) is not unique and it can be chosen such that
\be
(\langle 1|)\ \tilde{} = \langle 1 | ~~,~~(|0(\b_T) \rangle\ \tilde{} = |0(\b_T)\rangle.
\label{tildevacua}
\ee
These relations can be regarded as an operator representation of the KMS condition. The justification for taking these relations as axioms of the TFD method is that the vacuum expectation value between the above vacua corresponds to the thermal average as follows. By expressing $| 1 \rangle $ and $ |0(\b_T) \rangle $ as follows
\be
|1 \rangle = \sum_{n}|n\tilde{n}\rangle~~,~~|0(\b_T)\rangle = \frac{e^{-\b_T H}|1\rangle}{\mbox{Tr}[e^{-\b_T H}]},
\label{explicitthermalvacua}
\ee
where $|n\tilde{n}\rangle$ are the orthonormal basis vectors in the total Hilbert space, one obtains
\be
\langle 1 | A | 0(\b_T) \rangle = \frac{\mbox{Tr}[A \ e^{-\b_T H}]}{\mbox{Tr}[e^{-\b_T H}]}.
\label{fundamentalTFD}
\ee
One can note that $| 1 \rangle $ does not depend on the choice of the basis vectors and that it is a kind of identity state (hence the notation). The information about the thermal ground state is contained in the ket ground state 
$| 0(\b_T) \rangle$. The evolution of the thermal system is generated by the total Hamiltonian $\hat{H}$ which can be obtained from a total Lagrangian
\be
\hat{\LL} = \LL - \tilde{\LL}.
\label{totalLagrangian}
\ee 
In the canonical quantization, the formalism takes a simple operatorial form. For a linear oscillator the following relations hold
\bea
|1\rangle = e^{a^{\dagger}\tilde{a}^{\dagger}}|0\rangle,\label{bosonunit}
|1\rangle = (1 + i \ a^{\dagger}\tilde{a}^{\dagger})|0\rangle,\label{fermionunit}
\eea
for the bosonic and fermionic oscillators, respectively. Thus, the thermal bra and ket ground states can be taken symmetric and the formalism is the same as the formalism at zero temperature. The mapping from zero temperature to finite temperature is generated by the Bogoliubov operator
\be
G = - i\theta(\b_T)(a\tilde{a} - \tilde{a}^{\dagger}a^{\dagger}),
\label{Bogoliubovoscillator}
\ee
which is conserved
\be
[\hat{H},G] = 0.
\ee
In general, the Hilbert space at zero temperature and the thermal Hilbert space are not isomorphic for systems with an infinite number of degrees of freedom (see for further details \cite{ubook}.)

\section{The Bogoliubov Transformations of The Superstring Oscillators}

In order to apply the TFD to string theory, one has to modify the TFD ansatz in order to take into account the constraints. In the light cone gauge, this reduces to the inclusion of the level matching condition in the trace (see the Section 4 and the references \cite{ivv1,ng1,hemi}.)  

The Bogoliubov operator of a single oscillator can be given a linear form. In the case of superstrings, the
Bogoliubov transformations of the superstring oscillators act linearly on the oscillator operators as follows
\bea
a^{I}_n & \longrightarrow & a^{I}_n(\b_T ) = a^{I}_n \cosh \theta^{B,I}_n(\b_T ) - \tilde{a}^{I\dagger}_n \sinh \theta^{B,I}_n(\b_T ),\label{tra}\\
a^{I \dagger}_n & \longrightarrow & a^{I\dagger }_n(\b_T ) = a^{I \dagger}_n \cosh \theta^{B,I}_n(\b_T ) - \tilde{a}^{I}_n \sinh \theta^{B,I}_n(\b_T ),\label{tradagger}\\
\tilde{a}^{I}_n & \longrightarrow & \tilde{a}^{I}_n(\b_T ) = \tilde{a}^{I}_n \cosh \theta^{B,I}_n(\b_T ) - a^{I\dagger}_n \sinh \theta^{B,I}_n(\b_T ),\label{trtildea}\\
\tilde{a}^{I\dagger}_n & \longrightarrow & \tilde{a}^{I \dagger}_n(\b_T ) = \tilde{a}^{I \dagger}_n \cosh \theta^{B,I}_n(\b_T ) - a^{I}_n \sinh \theta^{B,I}_n(\b ),\label{trtildeadagger}\\
\bar{a}^{I}_n & \longrightarrow & \bar{a}^{I}_n(\b_T ) = \bar{a}^{I}_n \cosh \bar{\theta}^{B,I}_n(\b_T ) - \tilde{\bar{a}}^{I\dagger}_n \sinh \bar{\theta}^{B,I}_n(\b_T ),\label{bartra}\\
\bar{a}^{I \dagger}_n & \longrightarrow & \bar{a}^{I\dagger }_n(\b_T ) = \bar{a}^{I \dagger}_n \cosh \bar{\theta}^{B,I}_n(\b_T ) - \tilde{\bar{a}}^{I}_n \sinh \bar{\theta}^{B,I}_n(\b_T ),\label{bartradagger}\\
\tilde{\bar{a}}^{I}_n & \longrightarrow & \tilde{\bar{a}}^{I}_n(\b_T ) = \tilde{\bar{a}}^{I}_n \cosh \bar{\theta}^{B,I}_n(\b_T ) - \bar{a}^{I\dagger}_n \sinh \bar{\theta}^{B,I}_n(\b_T ),\label{bartrtildea}\\
\tilde{\bar{a}}^{I\dagger}_n & \longrightarrow & \tilde{\bar{a}}^{I \dagger}_n(\b_T ) = \tilde{a}^{I \dagger}_n \cosh \bar{\theta}^{B,I}_n(\b_T ) - \bar{a}^{I}_n \sinh \bar{\theta}^{B,I}_n(\b_T ),\label{bartrtildeadagger}\\
S^{a}_n &\longrightarrow & S^{a}_n (\b_T) = S^{a}_n \cos \theta^{F,a}_{n}(\b_T ) - \tilde{S}^{a\dagger}_n \sin \theta^{F,a}_{n}(\b_T ),\label{trS}\\
S^{a \dagger}_n &\longrightarrow & S^{a \dagger}_n (\b_T) = S^{a \dagger}_n \cos \theta^{F,a}_{n}(\b_T ) - \tilde{S}^{a}_n \sin \theta^{F,a}_{n}(\b_T ),\label{trSdagger}\\
\tilde{S}^{a}_n &\longrightarrow & \tilde{S}^{a}_n (\b_T) = \tilde{S}^{a}_n \cos \theta^{F,a}_{n}(\b_T ) + S^{a\dagger}_n \sin \theta^{F,a}_{n}(\b_T ),\label{trtildeS}\\
\tilde{S}^{a\dagger}_n &\longrightarrow & \tilde{S}^{a \dagger}_n (\b_T) = \tilde{S}^{a \dagger}_n \cos \theta^{F,a}_{n}(\b_T ) + \tilde{S}^{a}_n \sin \theta^{F,a}_{n}(\b_T ),\label{trtildeSdagger}\\
\bar{S}^{a}_n &\longrightarrow & \bar{S}^{a}_n (\b_T) = \bar{S}^{a}_n \cos \bar{\theta}^{F,a}_{n}(\b_T ) - \tilde{\bar{S}}^{a\dagger}_n \sin \bar{\theta}^{F,a}_{n}(\b_T ),\label{bartrS}\\
\bar{S}^{a \dagger}_n &\longrightarrow & \bar{S}^{a \dagger}_n (\b_T) = \bar{S}^{a \dagger}_n \cos \bar{\theta}^{F,a}_{n}(\b_T ) - \tilde{\bar{S}}^{a}_n \sin \bar{\theta}^{F,a}_{n}(\b_T ),\label{bartrSdagger}\\
\tilde{\bar{S}}^{a}_n &\longrightarrow & \tilde{\bar{S}}^{a}_n (\b_T) = \tilde{S}^{a}_n \cos \bar{\theta}^{F,a}_{n}(\b_T ) + \bar{S}^{a\dagger}_n \sin \bar{\theta}^{F,a}_{n}(\b_T ),\label{bartrtildeS}\\
\tilde{\bar{S}}^{a\dagger}_n &\longrightarrow & \tilde{\bar{S}}^{a \dagger}_n (\b_T) = \tilde{\bar{S}}^{a \dagger}_n \cos \bar{\theta}^{F,a}_{n}(\b_T ) + \tilde{\bar{S}}^{a}_n \sin \bar{\theta}^{F,a}_{n}(\b_T ).\label{bartrtildeSdagger}
\eea
Here, $\theta^{B,I}_{n}(\beta_T)$ and $\theta^{F,a}_{n}(\beta_T)$ are factors that depend on the temperature through $\beta_T = (k_B T)^{-1}$ and on the $n$-th oscillator frequency $\o_n$. Therefore, these factors will differ only for different oscillators $n$, but once $n$ is given, they are the same in all spacetime directions $I=1,\ldots ,8$ and for all spinor components $a=1,\ldots ,8$, respectively. Thus, the $I$ and $a$ indices can be dropped. Also, since the left-moving and right-moving oscillators of the same $n$ should be identical by symmetry between the left-moving and right-moving modes of the closed superstring, the $\theta_n$-factors should be the same in both left-moving and right-moving sectors and $\ \bar{} \ $ can be dropped, too. The form of $\theta$'s is given by the following relations \cite{ubook}
\be
\theta^{B}_{n}(\beta_T ) = \mbox{arccosh} (1-e^{-\beta_T \omega^{B}_n})^{-\frac{1}{2}}~,~
\theta^{F}_{n}(\beta_T ) = \mbox{arccos} (1+e^{-\beta_T \omega^{F}_n})^{-\frac{1}{2}},
\label{thetas}
\ee
where $\omega^{B}_n$ and $\omega^{F}_{n}$ are the frequencies of $n$-th bosonic and fermionic oscillators, respectively.
The generator of these transformations is just the algebraic sum of all Bogoliubov generators
\be
G = G^B + G^F,
\label{BogOp}
\ee
where $G^B$ and $G^F$ denote the sums of bosonic and fermionic Bogoliubov operators
\be
G^B = \sum_{n=1}^{\infty}\, (G^{B}_n + \bar{G}^{B}_n )~,~
G^F = \sum_{n=1}^{\infty}\, (G^{F}_n + \bar{G}^{F}_n ),
\label{BogOpn}
\ee  
and each term has the following form
\bea
G^{B}_n &=& -i\theta^{B}_{n}(\beta_T)\sum_{I=1}^{8}(a^{I}_n \, \tilde{a}^{I}_n - \tilde{a}^{I\dagger}_n \, a^{I \dagger}_n )\label{GBn},\\
\bar{G}^{B}_{n} &=& -i\theta^{B}_{n}(\beta_T)\sum_{I=1}^{8}(\bar{a}^{I}_n \, \tilde{\bar{a}}^{I}_n - \tilde{\bar{a}}^{I\dagger}_n \, \bar{a}^{I \dagger}_n )\label{barGBn},\\
G^{F}_{n} &=& -i\theta^{F}_{n}(\beta_T)\sum_{a=1}^{8}(S^{a}_n \, \tilde{S}^{a}_n - \tilde{S}^{a\dagger}_n \, 
S^{a \dagger}_n )\label{GFn},\\
\bar{G}^B &=& -i\theta^{F}_{n}(\beta_T)\sum_{a=1}^{8}(\bar{S}^{a}_n \, \tilde{\bar{S}}^{a}_n - \tilde{\bar{S}}^{a\dagger}_n \, \bar{S}^{a \dagger}_n )
\label{barGFn}.
\eea
The zero mode operators from the fermionic sectors $S^{a}_{0}$, $\bar{S}^{a}_{0}$, $\tilde{S}^{a}_{0}$ and $\tilde{\bar{S}}^{a}_{0}$ are isomorphic to the Dirac matrices. Therefore, we consider them inert under the thermalization. Another argument for that is that the $\theta(\b_T)$ parameter of the corresponding degrees of freedom, as defined by TFD, should be equal to $\pi /4$ which corresponds to an oscillator of zero energy. It follows that the Bogoliubov operator is made of operators of non-zero modes only and zero mode fermionic operators commute with the Bogoliubov operator.

The entropy of the superstring oscillators in $k_B$ units is given by the expectation value of the following operator in the thermal vacuum state \cite{ubook} 
\be
K = K_b + \bar{K}_b + K_{f} + \bar{K}_f,
\label{Entropyoperator}
\ee 
where 
\bea
K_b &=& - \sum_{I=1}^{8}\sum_{n=1}^{\infty}\left(a^{I\dagger}_n a^{I}_n\, \mbox{log}\,\mbox{sinh}^2\,\theta^{B}_n - a^{I}_n a^{I\dagger}_n\, \mbox{log}\,\mbox{cosh}^2\,\theta^{B}_n \right),\label{Kb1}\\
\bar{K}_b &=& - \sum_{I=1}^{8}\sum_{n=1}^{\infty}\left(\bar{a}^{I\dagger}_n \bar{a}^{I}_n\, \mbox{log}\,\mbox{sinh}^2\,\theta^{B}_n - \bar{a}^{I}_n \bar{a}^{I\dagger}_n\, \mbox{log}\,\mbox{cosh}^2\,\theta^{B}_n \right),\label{Kb2}\\
K_f &=& - \sum_{a=1}^{8}\sum_{n=1}^{\infty}\left(S^{a\dagger}_n S^{a}_n\, \mbox{log}\,\mbox{sin}^2\,\theta^{F}_n + S^{a}_n S^{a\dagger}_n\, \mbox{log}\,\mbox{cos}^2\,\theta^{F}_n \right),\label{Kf1}\\
\bar{K}_f &=& - \sum_{a=1}^{8}\sum_{n=1}^{\infty}\left(\bar{S}^{a\dagger}_n \bar{S}^{a}_n\, \mbox{log}\,\mbox{sin}^2\,\theta^{F}_n + \bar{S}^{a}_n \bar{S}^{a\dagger}_n\, \mbox{log}\,\mbox{cos}^2\,\theta^{F}_n \right).\label{Kf2}
\eea
The entropy operator takes into account only superstring oscillators at $T =0 $. Including operators from the tilde superstring would mean to take the average over the reservoire degrees of freedom, too.

\end{document}